\documentclass[epj]{webofc}
\usepackage[varg]{txfonts}   

\def\lsim{\mathrel{\rlap{\lower4pt\hbox{\hskip1pt$\sim$}}
    \raise1pt\hbox{$<$}}}
\def\gsim{\mathrel{\rlap{\lower4pt\hbox{\hskip1pt$\sim$}}
    \raise1pt\hbox{$>$}}}
\newlength{\dummysp}
\newcommand{\be}{\begin{equation}}
\newcommand{\ee}{\end{equation}}
\def\[{\left [}
\def\]{\right ]}
\def\({\left (}
\def\){\right )}

\wocname{EPJ Web of Conferences}
\woctitle{CONF16}
\begin{document}
\selectlanguage{english}
\title{Roundtable: What can we learn about confinement and anomalous
effects in QCD using analog systems? }
%
%

\author{M.~Cristina Diamantini\inst{1}\fnsep
             \thanks{\email{cristina.diamantini@pg.infn.it}} \and
        Dmitri Kharzeev\inst{2,3,4}\fnsep
              \thanks{\email{dmitri.kharzeev@stonybrook.edu}} \and
        Alexander Molochkov\inst{5} 
             \thanks{\email{molochkov.av@dvfu.ru}}  \and
        Thomas Sch\"afer\inst{6}\fnsep
              \thanks{\email{tmschaef@ncsu.edu}} \and
        Tin Sulejmanpa{\v s}i{\'c}\inst{7}\fnsep
              \thanks{\email{tin2019@gmail.com}}
}

\institute{NiPS Laboratory, INFN and Dipartimento di Fisica, 
           University of Perugia, via A. Pascoli, I-06100 Perugia, Italy.
\and
           Department of Physics and Astronomy, Stony Brook University, 
           NY 11794-3800 
\and
           Physics Department, Brookhaven National Laboratory, Upton, 
           NY 11973-5000 
\and
           RIKEN-BNL Research Center, Brookhaven National Laboratory, 
           Upton, NY 11973-5000  
\and
           Laboratory of Physics of Living Matter,
           Far Eastern Federal University, 
           Sukhanova 8, Vladivostok, 690950, Russia
\and
           Department of Physics,
           North Carolina State University, Raleigh, NC 27695
\and
           Philippe Meyer Institute, Physics Department,
           {\'E}cole Normale Sup{\'e}rieure, PSL Research University,
           24 rue Lhomond, F-75231 Paris Cedex 05, France
}

\abstract{%
  We discuss a number of examples for recent connections between 
emergent phenomena in many-body systems in atomic and condensed 
matter physics, and confinement and other non-perturbative 
effects in quantum chromodynamics.}
\maketitle
\section{Introduction}
\label{intro}

 In recent years, experimentalists in atomic and condensed matter physics
have achieved extraordinary control over ``designer'' many-body systems. 
This includes cold atomic Bose and Fermi gases with tuneable interactions
\cite{Bloch:2007}, atoms in optical lattices with controlled geometry
\cite{Bloch:2005}, and the design of artificial gauge fields 
\cite{Lewenstein:2006}. In condensed matter physics we have witnessed
the emergence of new classes of materials with designer Fermi surfaces, 
including Weyl and Dirac cones, and topologically protected surface states 
\cite{Qi:2011}. 

 This series of conference series is devoted to understanding strongly
correlated quantum field theories, in particular QCD, and to unraveling
the mechanism underlying emergent phenomena such as confinement and
chiral symmetry breaking. The question is whether the progress in 
atomic and condensed matter physics can contribute to these goals,
and whether some of things we have learned in QCD can have an impact
in other areas of physics. We can obviously not give a definitive 
answer or comprehensive summary of these issues here. Instead we
will discuss some examples in the following sections. We believe
that in general, there are several ways in which atomic and condensed
matter systems can have an impact:

\begin{itemize}
\item Universality is the observation that many phenomena are 
independent of the detailed microscopic dynamics. Historically,
universality has played a role in the study of continuous phase
transitions, and in effective field theory arguments. More recently, 
new ideas have emerged concerning universal aspects of anomalous
transport, and of universal transport in the strongly coupled regime, see
Sect.~\ref{sec-Thomas} and \ref{sec-Dima}. 

\item Duality expresses a weak-strong coupling or high-low temperature
equivalence between two different theories, typically formulated in 
terms of different degrees of freedom. While we may not be able to
construct a dual of (large $N$) QCD, dualities relevant to certain regimes 
of QCD can be studied in analog systems, see Sect.~\ref{sec-Cristina} 
and \ref{sec-Tin}.

\item Topology and semiclassical objects like instantons, monopoles, 
strings, and domain walls have long been discussed in QCD, but they
are not directly observable in the strong interaction. Analog systems
provide an opportunity to study topological objects in controlled 
settings, see Sect.~\ref{sec-Tin} and Sect.~\ref{sec-Dima},\ref{sec-Sasha}. 

\item Quantum simulators are the ultimate goal of many experiments
with trapped atoms or ions. The idea is to simulate the real
time evolution of an arbitrary Hamiltonian, including the evolution
of non-abelian gauge fields \cite{Wiese:2014rla}. At present a number 
of schemes for implementing abelian or non-abelian gauge fields have 
been tested in small systems, but the experiments typically do not 
realize dynamical gauge fields, and mostly explorer systems in reduced 
numbers of dimensions.

\end{itemize}

\begin{figure}[t!]
\begin{center}
\includegraphics*[width=15cm]{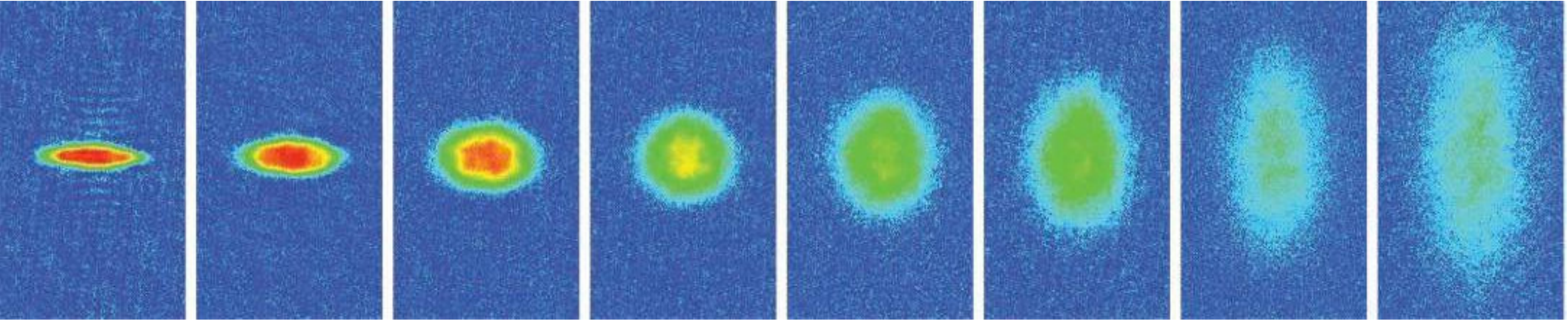}
\end{center}
\caption{\label{fig_cag_flow}
Expansion of a dilute Fermi gas at unitarity \cite{oHara:2002}.
The cloud contains $N\simeq 1.5\cdot 10^5$ $^6$Li atoms at a 
temperature $T\simeq 8\mu$K. 
The figure shows a series of false color absorption images taken 
between $t=(0.1-2.0)$ ms. The scale of the images is the same.
The axial size of the cloud remains nearly constant as the transverse
size is increasing. }
\end{figure}

\section{Thomas Sch\"afer: Transport Properties of Strongly Correlated
Fluids}
\label{sec-Thomas}

 Experiments at RHIC and the LHC have shown that quark gluon plasma
behaves as a nearly perfect fluid. This means that the shear viscosity
to entropy density ratio $\eta/s$ of the plasma is close to the limits
thought to be set by quantum mechanics \cite{Danielewicz:1984ww}.
Holographic dualities have provided a theoretical example of a
nearly perfect fluid \cite{Kovtun:2004de}, but we do not know whether
the physical mechanism for perfect fluidity in AdS/CFT is relevant 
to QCD. This motivates us to look for other strongly coupled 
fluids in nature, and to study whether a universal mechanism is 
at work \cite{Schafer:2009dj}.

 A very clean system that has been studied in some detail is
an ultracold atomic Fermi gas at unitarity. The system contains
two spin states of a neutral fermionic atom. The temperature and
density are so low that the interaction between atoms is fully 
characterized by the $s$-wave scattering amplitude. This scattering
amplitude can be tuned using a Feshbach resonance. The most 
interesting case is the limit in which the $s$-wave scattering
length is infinite, and the scattering amplitude is given by the 
universal unitarity limit, ${\cal A}=1/(ik)$, where $k$ is the 
scattering momentum. In this limit the system is very strongly
coupled, scale invariant, and universal (independent of the atomic
species). 

 Low viscosity fluid dynamics was first observed by O'Hara et
al.~\cite{oHara:2002}, see Fig.~\ref{fig_cag_flow}. The fluid
is prepared in a deformed trap and then released. Pressure gradients
in the fluid lead to acceleration, and because of the trap geometry
the acceleration is largest in the short direction. This phenomenon
is known as elliptic flow in the heavy ion community. In the case
of heavy ions elliptic deformation arises from the shape of the 
overlap region in a non-central collision. A remarkable aspect
of Fig.~\ref{fig_cag_flow} is that we can directly observe the
hydrodynamic evolution of the cloud. In the heavy ion case, 
the evolution has to be inferred from correlations among particles
in the final state. 

\begin{figure}[t!]
\begin{center}
\includegraphics*[width=7.cm]{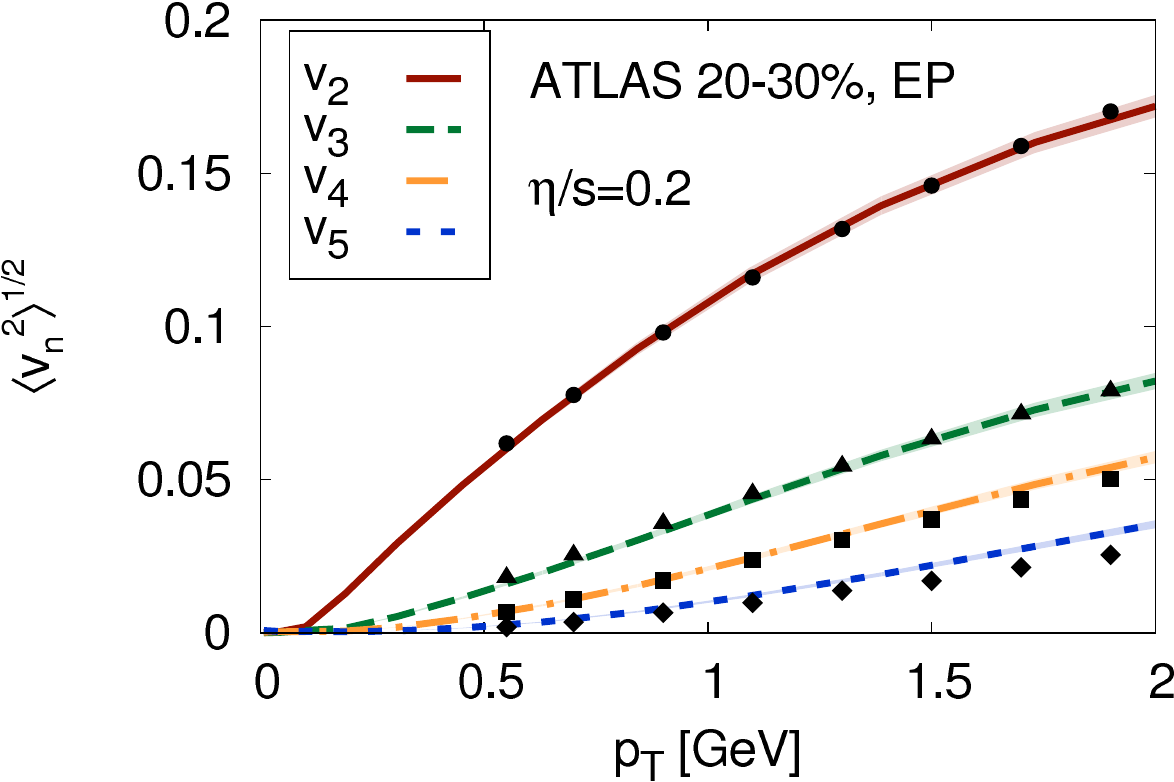}
\includegraphics*[width=7.cm]{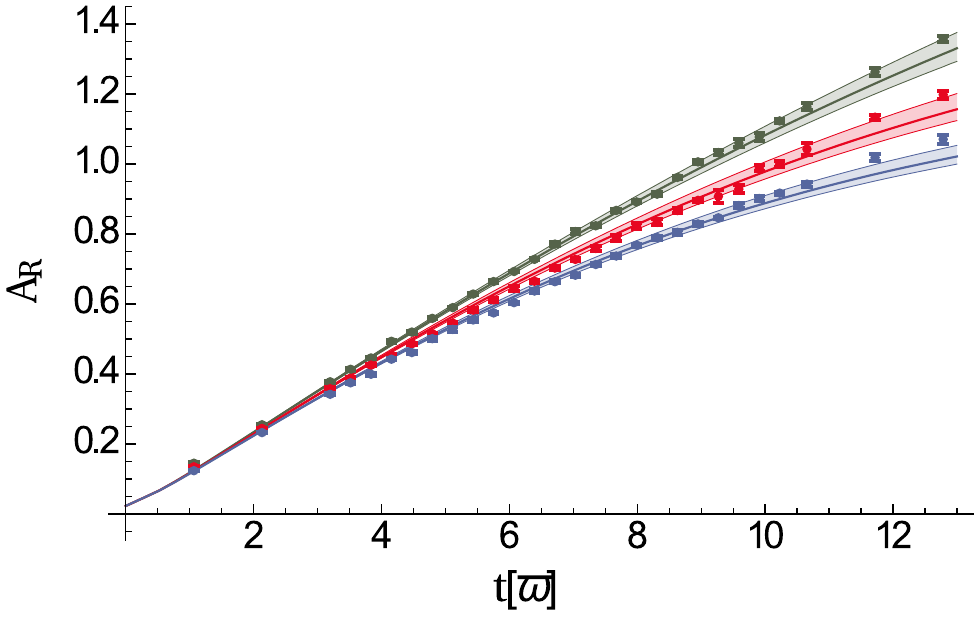}
\end{center}
\caption{\label{fig_vn}
Left panel: Fourier coefficients $v_2,\ldots,v_5$ of the azimuthal charged
particle distribution as a function of the transverse momentum
$p_T$ measured in $Pb+Pb$ collisions at the LHC \cite{ATLAS:2012at}
The lines show a hydrodynamic analysis performed using $\eta/s=0.2$
\cite{Gale:2012rq}. Right panel: Hydrodynamic fit to the expansion
data of Cao et al.~\cite{Cao:2010wa}, from \cite{Bluhm:2015bzi}. The 
data correspond to three different temperatures that are significantly 
above the critical temperature, and the fit was used to verify 
the validity of kinetic theory in the high temperature regime. }
\end{figure}

 Viscosity counteracts the differential expansion of the cloud, 
and detailed measurements of the final geometry provide constraints
on $\eta/s$. In the heavy ion case the initial geometry is not 
precisely known, and fluctuates from collision to collision. The 
best constraints arise from combining several harmonics of the flow
distribution, see the left panel of Fig.~\ref{fig_vn}. In the case 
of cold gases we can directly analyze the time evolution of the 
cloud shape, see the right panel of Fig.~\ref{fig_vn}. It is 
interesting to note that there is some ``technology transfer'' 
between heavy ions and cold atomic gases. The hydrodynamic fit
in the right panel was performed using a method called anisotropic 
fluid dynamics \cite{Bluhm:2015raa}, that was originally developed 
in the heavy ion context \cite{Florkowski:2010,Martinez:2010sc}. It 
is also interesting that in the cold gas context it is possible to
verify the validity of perturbation theory and kinetic theory
in the high temperature limit \cite{Bluhm:2015bzi}. This is not possible
in the case of heavy ions, both because the temperatures are 
not high enough, even at the LHC, and because at weak coupling
hydrodynamics is no longer useful. 

 The heavy ion data correspond to $\eta/s\simeq 0.2$. The current
best estimate for cold atomic gases near $T_c$ is $\eta/s\gsim 0.4$
\cite{Joseph:2014}. These two numbers are remarkably close, given
that the absolute magnitude of $\eta$ for the two systems differs by 
than 25 orders of magnitude \cite{Adams:2012th}. The question is whether 
this implies that there is a common mechanism at work. The mechanism
for nearly perfect fluidity in the quark gluon plasma has remained
somewhat elusive. Holographic methods account remarkably well for 
many properties of the plasma, like the short equilibration time. 
However, it is difficult to exclude that strongly interacting 
quasi-particles are present. In the cold atomic systems there
are a number of strategies for exploring the nature of perfect 
fluidity. These include measurement of the dynamic structure factor
\cite{Vale:2016}, numerical calculations of the viscosity spectral 
function \cite{Wlazlowski:2013owa}, and studies of non-hydrodynamic 
modes \cite{Brewer:2015ipa}. We expect these investigations to further 
enhance the interaction between the QCD, cold atomic, and string
theory communities.

\section{M.~Cristina Diamantini: Emergent versus reductionist approach}
\label{sec-Cristina}

The second law of thermodynamics and irreversibility are the paradigm of emergent phenomena that cannot be understood in terms of to elementary components but   as  collective phenomena.
In condensed matter, the idea of emergence in the interpretation of natural phenomena, as opposed to the reductionist approach, has led to many new important results.

The discovery  of the fractional quantum Hall  (FQHE) \cite{dqhl} effect has revealed the existence of a new state of 
matter characterized by a new type of quantum order, topological order \cite{wen1},
describing zero-temperature properties of a ground state with a gap for all excitations. 
Its hallmark are the degeneracy of the ground state on manifolds
with non-trivial topology. Although elementary constituents are electrons, the excitations over the incompressible ground state are anyons, quasi-particles  that exhibits fractional spin, statistics and  charge. If we think of the Laughlin state with filling fraction $\nu = 1/3 $, the idea is to "decompose"  the electron with charge e into 3 fractionally charged, fermionic  partons, each of them forming a $\nu =1$ integer quantum Hall effect. The strong interactions then "recombine"  the partons  to form a new emergent topological state of matter.
 
A characteristic feature of topological states is the existence of massless edge excitations that ensure  the stability of the topological phase: the bulk is gapped so the response to external perturbations  manifests itself through the edge dynamics. 
The edge excitations of chiral states, such as quantum Hall states, are stable,  protected  by the chiral anomaly: the non conservation of the charge at the boundary is compensated by the bulk current, a phenomenon know as anomaly inflow, and  leads to the quantization of the Hall conductance.
The situation is different for non-chiral topological states. 

Topological insulators are non chiral topological states, i.e. topological states that preserve parity and time reversal (TR) symmetry and exist in two and three dimensions.
For these materials, the edge excitations can become gapful, leading to  topologically trivial phases. 
 Symmetries, however, can in some cases  forbid the edge excitations from becoming gapped.
Topological insulators , e.g., are protected by TR symmetry and are an example of symmetry-protected topological phases of matter \cite{chen}. 
In the case of 3-dimensional topological insulators this leads to the presence of a topological $\theta$-term in the bulk action:
\begin{equation}
S  = \int d^4x {i  \theta  \over 16 \pi^2} F_{\mu \nu}\tilde F^{\mu \nu} 
 = \int d^4x {i  \theta  \over 4  \pi^2} {\bf E} \cdot {\bf B} \  ,
\end{equation} 
where $F_{\mu \nu}$ is the Maxwell field strength, $\tilde F^{\mu \nu}$ its dual, and  $\theta = 0 $ for trivial topological insulators and $\theta = \pi$ for strong topological insulators.

An important characteristic of topological states of matter is that they have  a confinement phase \cite{noi}   described by an action that is the same as the confining string action proposed by Polyakov  and Quevedo and Trugenberger \cite{pol} to describe confinement in compact QED, and used as a toy model to understand confinement in QCD.
In fact topological matter is characterized by the presence of a topological BF term in its long-distance effective action \cite{noi}. In three dimensions the $U(1)$ symmetry associated with the BF action is compact. Topological defects due to the compactness of the $U(1)$ gauge fields induce quantum phase transitions between topological insulators, topological superconductors (TS) and topological confinement (TC). In conventional superconductivity, the photon acquires a mass due to the Anderson-Higgs mechanism. In TS, instead,  magnetic flux is confined and the photon acquires a topological mass through the BF mechanism \cite{noi}: no symmetry breaking is involved, the ground state has topological order and the transition is induced by quantum fluctuations. There is no Higgs fields, the degree of freedom that is "eaten" by the photon is a scalar mode due to the condensation of electric solitons.

The TC phase is dual to topological superconductivity: in this case it is the new degree of freedom arising from the condensation of magnetic vortex strings that "eats" the original photon via the Stueckelberg mechanism. This mechanism turns the photon into a massive antisymmetric tensor that couples to electric strings between charge-anticharge pairs, thus realizing linear charge confinement.
The equation that a describes this phase is the confining string action:
\begin{equation}
S_{\rm eff}^{TC} = 
 +\int d^4x  {1\over 4} B_{\mu \nu}B_{\mu \nu}
 + {1\over 12\Lambda ^2} H_{\mu \nu \alpha} H_{\mu \nu \alpha} 
  + B_{\mu \nu} T_{\mu \nu}  \ ,
\end{equation}
where $B_{\mu \nu}$ is the Kalb -Ramond field, $H_{\mu \nu\alpha}$ its field strength. $T_{\mu \nu} $ parametrises the world-sheet of the string.

Can we use topological phases of matter as toy models to understand QCD? In \cite{laugh1}, Laughlin argues that confinement can be interpreted as a collective phenomenon, relating the interpretation of confinement with the work of Senthil \cite{sen} on deconfined quantum criticality, according to which fractional quantum numbers emerge at  phase transitions and become confined away from criticality due to strong interactions, as it happens for topological states of matter.

The idea that QCD can be interpreted as topologically ordered phase as been investigated in \cite{zit} with an analysis based on "deformed" QCD. In \cite{tac} the analogy between the structure QCD vacuum and topological insulators is investigated, with emphasis on the role of the $\theta$-term in both cases.

\section{Tin Sulejmanpa{\v s}i{\'c}: Liberation on domain walls in gauge 
theories and quantum magnets}
\label{sec-Tin}

\begin{figure}[t] 
\centering
\includegraphics[width=.50\textwidth]{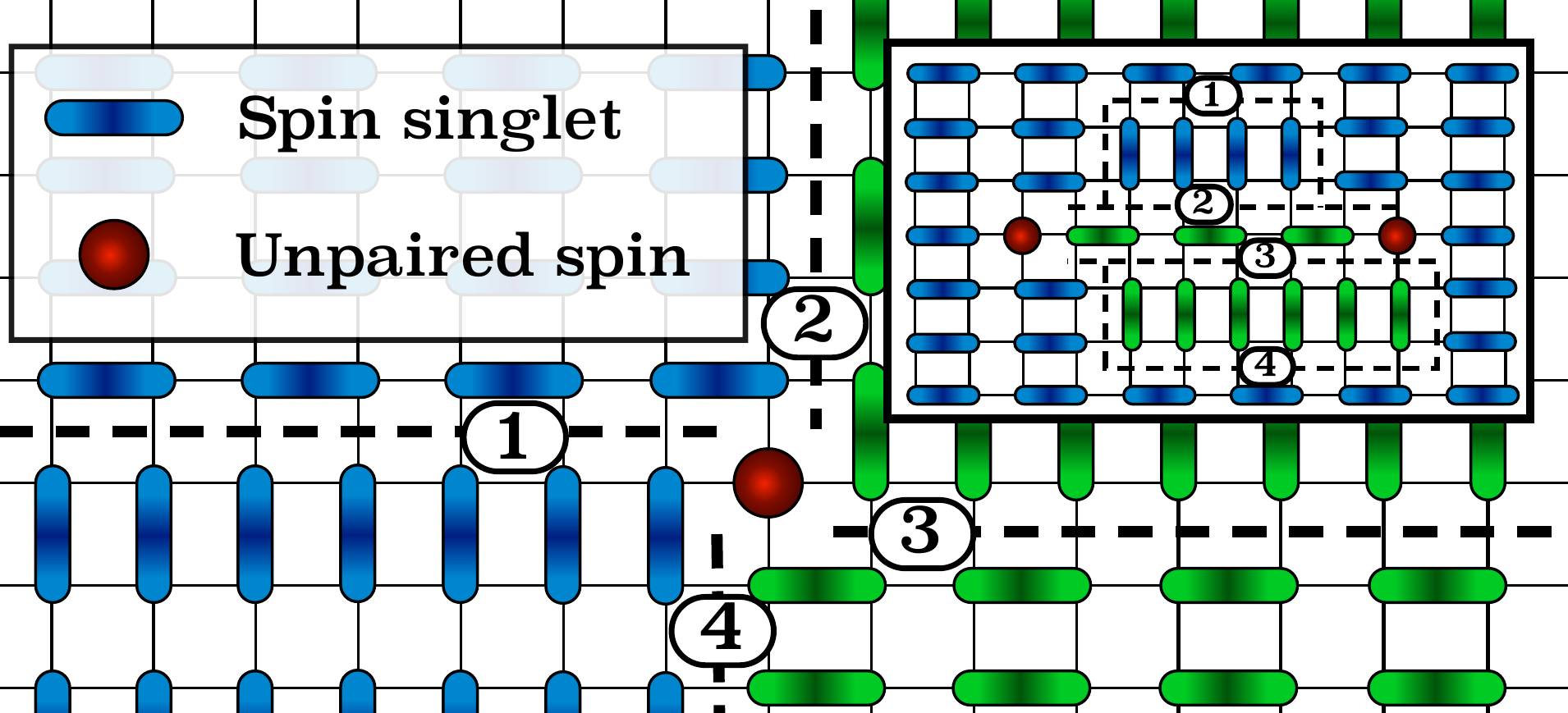} 
\caption{A sketch of the four ground state of the VBS in 2D on a cubic 
lattice. The green and blue rods indicate spin singlet states, the dashed 
lines indicate domain walls separating the different vacua, while at the 
nexus of the picture an unpaired spin results from the intersection of the 
four domain walls. The inlay shows a sketch of the composite nature of the 
confining strings.}
\label{fig:DW_main}
\end{figure}

Confinement, although well known in QCD, is not exclusive to it. Similar behavior occurs in quantum anti-ferromagnets in the so-called Valence Bond Solid (VBS) phase. This state is most easily described in the basis where spin states on lattice sites are paired into singlets, i.e. dimers. The ground state of the VBS can be visualized as crystal ordering of these dimers, spontaneously breaking lattice symmetries. In the case of two spatial dimensions (2D) on the cubic lattice, the $Z_4$ lattice rotational symmetry is completely  spontaneously broken by the long range order of dimers, ensuing four degenerate vacua (see Fig. \ref{fig:DW_main}). Such vacua must be separated by domain-walls (or lines in the case of 2D lattice). It is obvious from Fig. \ref{fig:DW_main} that if these vacua are put side by side, at their nexus there must be an unpaired spin -- a spinon \cite{Levin:2004}. In other words, a spinon excitation \emph{sources four domain walls}. Confinement of spinons is manifest in the fact that these domain-walls carry energy proportional to their length. A single spinon excitation would therefore carry an infinite energy in the thermodynamical limit and would not appear in the physical spectrum. In other words a pair of spinons would be bound by four domain-wall strings\footnote{Keep in mind that the string will inevitably break at some point because the spinons are inherently dynamical.}. Weakening the VBS order eventually causes a transition to the Ne\'el ordered state (anti-ferromagnetic order). This transition has been of interest in the study of so-called deconfined-quantum criticality \cite{senthil2004,senthil2004b}.

So what does this, if anything, have to do with gauge theories? Naively very little. In VBS confinement is a topological phenomena: an unpaired spinon misaligns the singlet dimers and forces interfaces between the degenerate vacua. Such interfaces cost a fixed energy per unit length, and must be terminated by another spinon in order to avoid paying infinite energy. In QCD there are not even degenerate vacua, so naively this picture of confinement has little to do with QCD or Yang-Mills theory. 

However it was known for a long time that while Yang-Mills theories do not have multiple \emph{degenerate} vacua, large $N$ arguments indicate that it does have non-degenerate k-vacua, which were first conjectured by Witten \cite{Witten:1980sp}. Witten argued that, on the one hand, the 't Hooft large-$N$ limit forces the vacuum energy to be a function of $\theta/N$. On the other hand the partition function must be a periodic function of the $\theta$-angle. The resolution is that in the large $N$ limit the vacuum energy is given by
\be
E(\theta)=N^2 \min_{k}f((\theta-2\pi k)/N)\;,\;\; k\in \mathbb Z\;,
\ee
in other words the energy as a function of the $\theta$-angle is a multi-branch function, where branches are labeled by the integer $k$.
For $\theta\in[0,\pi)$ the value of $k$ is taken to be zero. When $\theta=\pi$ the vacua $k=0$ and $k=1$ are degenerate and are related by the $CP$ symmetry which is (believed to be) spontaneously broken in the confined phase. Further if massless matter fields in an adjoint gauge-group representation are added to the pure theory, the theory is believed to spontaneously break the discrete $Z_{2N}$ chiral symmetry\footnote{See for example the discussion in \cite{Unsal:2007jx}} down to $Z_2$, ensuing $N$-discrete vacua.  Could confinement in such theories have any connection with the confinement in the VBS of quantum anti-ferromagnets? 

In fact in 2007 a novel way to study the QCD-like theories was proposed by M. \"Unsal et. al. \cite{Unsal:2007jx,Unsal:2008ch,Shifman:2008ja}. Motivated in part by the Eguchi-Kawai ideas, \"Unsal proposed that one space-time direction of the theory can be compactified on a small circle in such a way that no deconfinement phase transition occurs. A short distance scale of the compact circle controls the coupling of the theory, and as long as the compact circle is small enough, the theory is weakly coupled. Because of this, semi-classical methods can be used to understand the full structure of the confining theory.

\begin{figure}[t] 
\centering
\includegraphics[width=.50\textwidth]{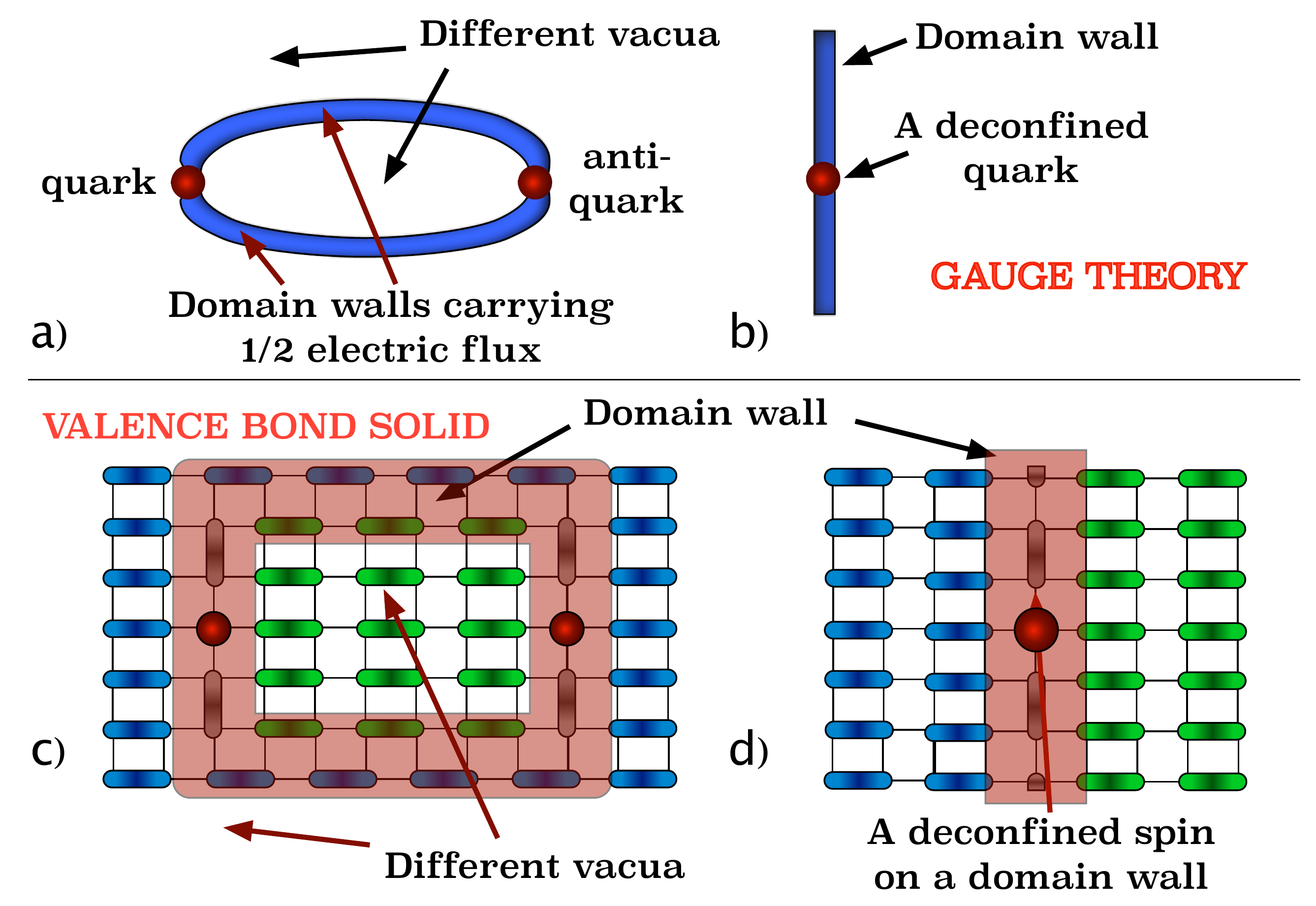}
\raisebox{1.1cm}{\includegraphics[width=.40\textwidth]{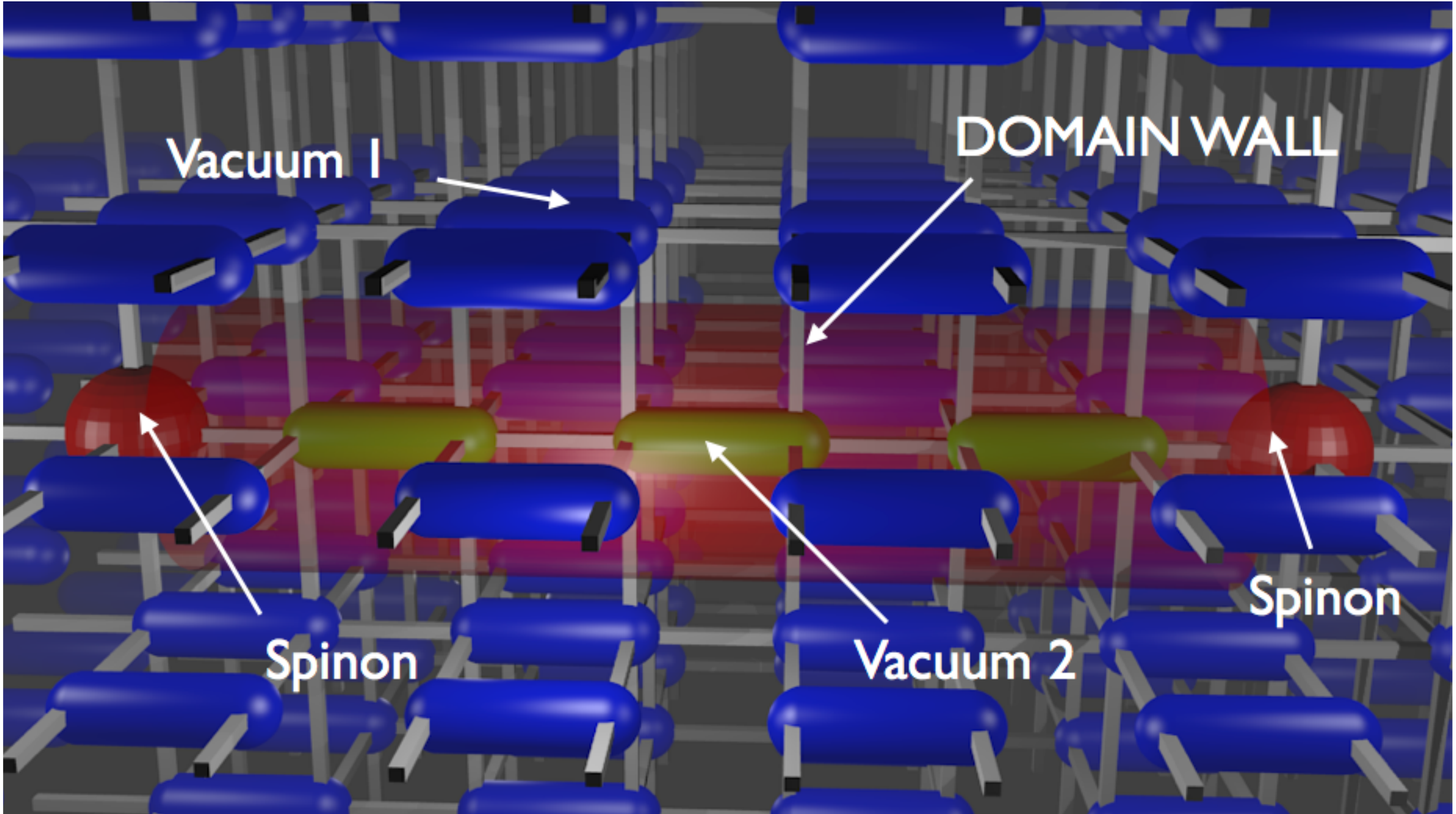}}
\caption{Left: A comparison of the quark/spinon confinement and wall-liberation 
in compactified QCD-like theories a), b) and of spinons in the VBS phase c), 
d). Right: 3D VBS string with $Z_2$ vacua. The red region indicates a 
domain-wall, i.e. the string is really a tubular domain wall.}
\label{fig:spins_and_quarks}
\end{figure}

Indeed using such reasoning Anber, Poppitz and TS \cite{Anber:2015kea} noticed that the confining mechanism in theories where there are multiple degenerate vacua in such semi-classical regimes does not conform to the standard picture of confinement one usually has in mind. Instead of a solid confining string, the confining strings are composed out of two domain-walls of the theory (see sketch in Fig.~\ref{fig:spins_and_quarks}a), a picture strikingly similar to the one in VBS of quantum anti-ferromagnets discussed above\footnote{Shortly after \cite{Anber:2015kea}, similar string structure was noticed in the dimer model on a square lattice \cite{Banerjee:2015pnt}.}. The only difference is that the number of domain walls emanating from the spinon is four, not two.

In fact the similarity is no accident, and is a consequence of the fact that the underlying confining mechanisms in both gauge theories and in quantum anti-ferromagnets are strikingly similar: they result from the existence and influence of monopole-like instantons as explained in \cite{Sulejmanpasic:2016uwq}. So the naively vastly different confining regimes are more similar than one would naively think.

But this is not all. In gauge theories such a picture explained an old conjecture made in the context of super Yang-Mills theory by Soo Jong Rey \cite{Witten:1997ep}. S.J.~Rey argued from the microscopic picture of the Seiberg-Witten theory that the domain walls host deconfined fundamental quarks. In other words on the domain walls the hadrons fractionalize into their constituents. How this happens in the setup of \cite{Anber:2015kea} is simple to see from Fig. \ref{fig:spins_and_quarks} a) and b). By taking one of the quarks to infinity in Fig. \ref{fig:spins_and_quarks}a, a string made of domain walls can be deformed to Fig. \ref{fig:spins_and_quarks}b. A domain wall, literally being the makings of the string, completely absorbs all the flux emanating from the quark which is now free to move along the domain wall.

Can this phenomena be present in VBS? A bit of thought will quickly convince the reader that a domain wall of the VBS depicted in Fig. \ref{fig:DW_main} will not host deconfined spinons, as the domain wall is only able to absorb half of the ``confining string''. However should lattice symmetry be reduced to $Z_2$, a vastly different picture on the domain wall emerges. Spinons, now confined by two domain walls (see Fig. \ref{fig:spins_and_quarks}c) are liberated on walls and appear as physical particles there, as obvious from Fig. \ref{fig:spins_and_quarks}. An ideal model in which to test this numerically is the so-called J-Q model. The J-Q model on a cubic lattice is well known to be in the VBS phase for a certain range of its parameters, and it has been extensively used for numerical studies of the VBS phase and quantum critical phenomena in quantum magnets \cite{Sandvik:2006fpf,Tang:2013bza,Shao2015,Shao2016} (see \cite{Sandvik_review} for a review). Indeed as Shao, Sandvik, \"Unsal and TS demonstrated in \cite{Sulejmanpasic:2016uwq} that while the usual VBS typically has exponential correlation functions, in the presence of the domain wall, the $Z_2$ VBS shows algebraic correlations along the wall and is in perfect agreement with the 1D spin chain, which is well known to host spin $1/2$ excitations (spinons) \cite{shastry1981,faddeev1981,tennant1993}. This is in direct contrast to the excitations of the bulk VBS which are spin $1$ triplets.

This realization not only demonstrated a novel phenomena in quantum magnets, which may be directly observable in experiments and perhaps useful for the field of spintronics, but also made a direct connection to the non-abelian gauge field theory. One obvious question is how much can we learn from QCD about quantum magnets and vice versa. While the version of QCD realized in nature, as already emphasized, does not have degenerate vacua, it may still have non-degenerate $k$-vacua. In fact it is tempting to think that a part of the string tension is due to the excitation of the $k$-vacua. This would imply that under the deformation of the theory which render the vacua degenerate, the string tension should drop. Further, it is possible that the phenomena of deconfinement on the domain walls may be connected with the appearance of the edge modes in topological insulators. Should this analogy be correct it may yield a unique insight into the vacuum of QCD. Finally, similar numerical studies of higher dimensional VBS probably has a richer structure, and may possibly host a spin liquid which has been of interest in the physics of high $T_c$ superconductivity. Such systems may further realize confining strings as tubes of domain walls (see Fig. \ref{fig:spins_and_quarks} right), at least in a limit of parameters where the string thickness is comparable to the lattice size. Whether this scenario is realized in the continuum description, or whether it bears any relation to non-abelian gauge theories is at the moment unclear.  

\section{D.~Kharzeev: Chiral matter in QCD and condensed matter physics}
\label{sec-Dima}

The presence of nearly massless chiral fermions in the action of Quantum Chromodynamics is crucial for understanding the key properties of the theory. The chiral symmetry and its spontaneous breaking, as well as the properties of confinement in real world, result from the dynamics of strongly coupled light quarks. In particular, one may expect that the chiral anomaly assumes an important role by connecting the behavior of the theory to the global topology of gauge fields -- for example, it has been argued recently that ${\rm U_A(1)}$ anomaly may provide a mechanism for confinement \cite{Kharzeev:2015xsa}.    

Theoretical tools for understanding this dynamics in QCD are mostly limited to the chiral perturbation theory (in the spontaneously broken phase populated by Goldstone bosons), and lattice QCD in Euclidean space-time. It is thus desirable to have an experimental access to controllable systems possessing chiral fermions. 

In high energy nuclear physics, such an access is provided by relativistic heavy ion collisions creating hot and dense QCD matter. Moreover, it appears that the colliding ions create a very intense magnetic field \cite{Kharzeev:2007jp}, and so at least at early moments the dynamics of chiral quarks is affected both by strong and electromagnetic interactions. In this case, one can test a specific consequence of chiral anomaly for the transport of electric charge -- the chiral magnetic effect (CME) \cite{Fukushima:2008xe}, see 
\cite{Kharzeev:2013ffa,Kharzeev:2015znc} for reviews and additional references. Evidence for the CME and related phenomena has been reported by STAR Collaboration at RHIC \cite{Abelev:2009ac,Abelev:2009ad,Adamczyk:2015eqo} and ALICE Collaboration at the LHC \cite{Adam:2015vje}.  Reaching an unambiguous conclusion on the existence of CME in heavy ion collisions however is a challenging task, as there are substantial backgrounds. It should be possible to quantify them further using the data on proton-nucleus collisions recently presented by the CMS Collaboration at the LHC \cite{Khachatryan:2016got}, and related analyses at RHIC. The dedicated 2018 isobar  run at RHIC (using the nuclei with the same mass number but different electric charge) will allow to clearly distinguished between the background effects driven by nuclear geometry and the effects of magnetic field \cite{Skokov:2016yrj}. 

In the meantime, it is desirable to study the real-time dynamics of quantum chiral matter in other settings that may allow better control. It has been known for some time that in condensed matter systems the symmetries of the crystalline lattice sometimes allow the emergent chiral quasi-particles. A remarkable example of such a system is graphene  \cite{GN2007}, where the left-right symmetry of A and B sublattices of the hexagonal lattice translates into the emergence of massless chiral fermions in the low-energy spectrum. The role of the effective coupling constant in graphene is played by ${\bar \alpha} = e^2/\hbar v_F$, where the speed of light $c$ is substituted by the Fermi velocity of quasiparticles near the Dirac point. Experimentally, $v_F \simeq c/300$, so ${\bar \alpha} \simeq 2$ and we deal with strongly coupled fermions -- in fact, the coupling is as strong as the QCD coupling near the deconfinement phase transition! This suggests that the transport in graphene may resemble that of viscous fluids (such as the quark-gluon plasma) \cite{MSF}; in particular, an external electric field creates vortices that can drive electric current against an applied field \cite{LF2016}-- this opens the possibility to measure the viscosity of the strongly coupled fluid in experiment. It may also be possible to study in graphene the spontaneous breaking of chiral symmetry  \cite{Aleiner:2007va} -- in the language of condensed matter physics, the chiral condensate corresponds to the formation of the excitonic condensate. While this does not usually happen, the spontaneous symmetry breaking can be induced by applying an external magnetic field in the plane of graphene \cite{Aleiner:2007va}.

Graphene is a two-dimensional (or $2+1$ D, in the nomenclature of relativistic field theory) system, and so does not allow to study the effects of chiral anomaly -- it is well known that quantum anomalies exist only in even number of space-time dimensions.  However, the recent discovery of 3D chiral materials -- Dirac and Weyl semimetals -- opens exciting possibilities for the study of anomaly-induced transport in tabletop experiments. In particular, the chiral magnetic effect has been decisively observed in a Dirac semimetal ${\rm ZrTe_5}$ \cite{Li:2014bha}. The recent surge of experimental results on the CME and related phenomena in Dirac and Weyl semimetals points to the possibility to uncover the role of quantum effects in chiral matter, with important implications both for condensed matter physics and QCD. 

\section{A.~Molochkov: Gauge symmetries and structure of proteins}
\label{sec-Sasha}

One of the most unusual and interesting applications of the effective gauge field theory is the protein structure study.
  
Currently, the most ambitious computational approaches to modeling the 
structure of proteins based on classical molecular dynamics allow 
us to describe the processes of protein folding in the case of short and 
fast folding protein only~\cite{Feddolino}. For a realistic description 
of the tertiary and quaternary structures on large spatial and temporal 
scales is needed computing power by 5-6 orders of magnitude greater than 
is technically achievable now~\cite{Lindor}. 

One of the solutions of the problem can be obtained with the help of 
coarse grain modeling within effective field theory, which allows a 
natural way to introduce the collective degree of freedom and nonlinear 
topological structures based on fundamental principles of gauge symmetry. 
The corresponding field theory model is based on local symmetry of proteins 
that will dynamical define tertiary structure of proteins.
   
Protein local symmetry is defined by amino-acids and protein backbone
bond structure. The $C_{\alpha}$ backbone bonds are almost freely rotating, 
which leads to a $U(1)$ local symmetry. On the other hand hydrophobic and 
hydrophilic forces lead to the rotation symmetry breaking due to amino 
acid alignment inside (hydrophobic) or outside (hydrophilic) of the 
secondary or tertiary regular structure. 
Another phenomenologically known protein symmetry is chiral symmetry, which 
is broken due to chirality of amino acids. It leads to chirality of ground 
state structures like $\alpha$-helices.  

 A field theory model can be formulated within an approach that uses 
the description of the local geometry of proteins based on the formalism 
of discrete coordinates of Frenet~\cite{Niemi1}. Within this formalism 
proteins are considered as one-dimensional discrete units, which determine 
the free energy functional, defined solely by the angles of curvature and 
torsion. The Frenet Frame rotation can be presented by the following 
transformation: 
\begin{eqnarray}
\frac{d}{ds}\left(\begin{array}{c} 
{\bf e}_1 \\ {\bf e}_2\\{\bf t}\end{array} \right) 
= \left(\begin{array}{ccc} 
0 & (\tau+\eta_s) & -\kappa \cos(\eta)\\ 
    (\tau+\eta_s) & 0 & \kappa \sin(\eta) \\ 
  \kappa \cos(\eta) & -\kappa \sin(\eta)& 0
\end{array}\right)\times 
 \left( \begin{array}{c}{\bf e}_1 \\ {\bf e}_2\\{\bf t} \end{array} \right)
\end{eqnarray}
Under rotation of the local coordinate system the doublet of dynamic variables 
is transformed just as the two-dimensional Abelian Higgs multiplet. This 
transformation can be rewritten as a $U(1)$ gauge  transformation of the 
scalar field:
\begin{eqnarray} 
&\kappa \sim   \phi  \to \kappa e^{-i\eta}\equiv \phi e^{-i\eta}\\
&\tau   \sim  A_i    \to \tau + \eta_s \equiv A_i+ \eta_s	
\end{eqnarray}
Here the bending ($\kappa \sim \phi $) and torsion ($\tau \sim A_i$) are 
introduced as scalar and gauge filed correspondingly. Thus, the 
Abelian Higgs Model Hamiltonian takes the form:
\begin{equation} 
H=\int\limits_0^L ds\, \left(|(\partial_s+ie\tau)\kappa|^2
   +\lambda(|\kappa|^2-m^2)^2+a\tau\right)	
\label{Hmlt}
\end{equation}
term $a\tau$ is the Chern-Simons term ensuring chirality breaking - domination 
of the right hand alpha-helices. As the results we obtain ground states of 
the protein backbone as the theory vacua with different topological sectors: 
$\alpha$-helices with broken chiral symmetry and negative parity ($\kappa
\simeq \frac{\pi}{2}$, $\tau\simeq 1$), $\beta$-strands with restored 
chiral symmetry and positive parity ($\kappa\simeq 1$, $\tau\simeq \pi$).

Spatial transitions between different ground states are defined by the 
exact solutions of the real part of the Hamiltonian (\ref{Hmlt}) are 
nonlinear structure in the protein backbone bending like kinks~\cite{Niemi2}:
 \begin{equation}
\kappa(s)=m \tanh\left(m\sqrt{\lambda}(s-s_0)\right)
 \end{equation}
with energy:
 \begin{equation}
 E=\int ds \left(\kappa_s^2+\lambda(\kappa^2-m^2)^2
   -a\tau-b\kappa^2\tau+\frac{c}{2}\tau^2+\frac{d}{2}\kappa^2\tau^2\right).
 \end{equation}
Within this approach the structure of the protein backbone is parametrised 
by superpositions of the one-dimensional solitons (kinks). 

It allow us to use full power of the field theory approach to study not 
only protein structure in terms of the collective degrees of freedom, but 
also dynamics at different thermodynamical conditions and in external fields. 
For example, Glauber heating and cooling analysis within the presented model 
shows that the myoglobin unfolding to the molten globule and folding back 
occurs due to unfolding and folding of the F-helix, which is bound with the 
heme-group.   

\vskip0.3cm

Acknowledgments: This work was supported in parts by the US Department 
of Energy grant DE-FG02-03ER41260 (TS), the Federal Target Program
for Research and Development in Priority Areas of Development of the 
Russian Scientific and Technological Complex for 2014-2020, Contract 
14.584.21.0017 (AM), and the US Department of Energy under Contracts 
No. DE-FG- 88ER40388 and DE-AC02-98CH10886 (DK).

%
%
%

\end{document}